\documentclass{aastex}
\usepackage{emulateapj5,apjfonts,epsfig}

\shorttitle{ULTRACOMPACT X-RAY BINARIES}
\shortauthors{JUETT ET AL.}
\slugcomment{Accepted for publication in ApJ Letters}

\begin{document}
\bibliographystyle{apj_noskip}

\title{Ultracompact X-ray Binaries with Neon-Rich Degenerate Donors}

\author{Adrienne~M.~Juett, Dimitrios~Psaltis,\altaffilmark{1} and
        Deepto~Chakrabarty} 
\affil{\footnotesize Department of Physics and Center for Space Research,
	Massachusetts Institute of Technology, Cambridge, MA 02139; \\ 
	ajuett, demetris, deepto@space.mit.edu}

\altaffiltext{1}{Current address: School of Natural Sciences,
Institute for Advanced Study, Princeton, NJ 08540}

\begin{abstract}

There are three low-mass X-ray binaries (4U~0614+091, 2S~0918$-$549,
and 4U~1543$-$624) for which broad line emission near 0.7~keV was
previously reported.  A recent high-resolution observation of
4U~0614+091 with the {\em Chandra}/LETGS found evidence for an
unusually high Ne/O abundance ratio along the line of sight but failed
to detect the previously reported 0.7~keV feature.  We have made a
search of archival {\em ASCA\/} spectra and identified a fourth source
with the 0.7~keV feature, the 20-min ultracompact binary
4U~1850$-$087.  In all four of these sources, the 0.7~keV residual is
eliminated from the {\em ASCA\/} spectra by allowing excess
photoelectric absorption due to a non-solar relative abundance of
neon, just as in the LETGS spectrum of 4U~0614+091.  The optical
properties of these systems suggest that all four are ultracompact
($P_{\rm orb}<80$ min) binaries.  We propose that there is excess neon
local to each of these sources, as also found in the ultracompact
binary pulsar 4U~1626$-$67.  We suggest that the mass donor in these
systems is a low-mass, neon-rich degenerate dwarf and that the
binaries are all ultracompact.
\end{abstract}


\section{Introduction}

There have been many searches for line-like features in the
low-resolution X-ray spectra of neutron stars in low-mass X-ray
binaries (NS/LMXBs; see White, Nagase, \& Parmar 1995 for a review).
The detections fall into several categories.  A broad emission feature
near 0.7~keV has been repeatedly reported from 4U~0614+091 (Christian,
White, \& Swank 1994; White, Kallman, \& Angelini 1997; Schulz 1999;
Piraino et al. 1999).  A similar feature has also been reported from
2S~0918$-$549 and 4U~1543$-$624 (White et al. 1997).  Line-like
features near 1~keV have been reported from Cyg~X-2 and Sco~X-1
(Kuulkers et al. 1997; Vrtilek et al. 1991), and an unusual complex of
Ne/O emission lines is detected in the ultracompact LMXB pulsar
4U~1626$-$67 (Angelini et al. 1995; Owens, Oosterbroek, \& Parmar
1997; Schulz et al. 2001).  Fluorescent Fe $K$ line emission (6.4--6.7
keV) has been reported from a number of systems (e.g., Asai et
al. 2000).  More recently, high-resolution observations with the {\em
Chandra X-Ray Observatory} and {\em XMM-Newton} have led to a number
of new line detections (Brandt \& Schulz 2000; Cottam et al. 2001a,
2001b; Schulz et al. 2001).

In this paper, we focus on the origin of the 0.7~keV feature reported
from several sources.  A recent high-resolution observation of the
brightest of these, 4U~0614+091, with the {\em Chandra} Low Energy
Transmission Grating Spectrometer failed to detect the feature, a
failure attributed to source variability (Paerels et al. 2001).
Paerels et al. (2001) did however find an unusually high Ne/O 
abundance ratio from absorption edges, which we show could produce a 
line-like residual near 0.7~keV in low-resolution data when not taken 
into account.  We report here on a
comprehensive search of the LMXB spectra in the {\em ASCA\/} public
data archive for additional examples of sources with the 0.7~keV
feature.  We identify a fourth source with this feature, the
ultracompact LMXB 4U~1850$-$087.  Finally, we show that the {\em
ASCA\/} spectra of these four systems, as well as the {\em
Chandra}/LETGS spectrum of 4U~0614+091, can all be well fitted {\em
without\/} a 0.7~keV line feature if excess photoelectric absorption
due to neon-rich material near the source is present.

\section{Observations and Data Reduction}

The {\it ASCA\/} satellite (Tanaka, Inoue, \& Holt 1994) operated from
1993 February to 2000 July carrying four instruments: two Solid-State
Imaging Spectrometers (SIS0 and SIS1; 0.5--10.0~keV) and two Gas
Imaging Spectrometers (GIS2 and GIS3; 1.0--10.0~keV).  The SIS
instruments had superior resolution (0.07 keV at 1 keV), making them
better suited for soft X-ray line studies.  We used both {\tt BRIGHT}
and {\tt FAINT} mode SIS data in our analysis.  The SIS CCDs gradually
degraded after launch due to radiation damage, and we note that the
{\tt BRIGHT} mode data are more susceptible to this damage than the
{\tt FAINT} mode data.  However, since the observations we used all
took place before 1996, we believe that any systematic problems due to
radiation damage are smaller than the statistical uncertainties in the
data.  We used the XSPEC v11 spectral fitting package to perform the
analysis presented here (Arnaud 1996).

We examined the observations of all 56 NS/LMXBs in the {\it ASCA\/}
public data archive as of 2000 April in order to identify sources with
a 0.7~keV feature similar to that reported in 4U~0614$+$091.  For our
preliminary search, we used the preprocessed standard spectral data
products for each source and fit the SIS data with an absorbed
power-law or power-law $+$ blackbody model.  Nine sources showed
significant residual excess below 1~keV.  However, three of these
sources (4U~1254$-$690, 4U~1755$-$338, and 4U~2127$+$119) are known
X-ray dippers and were excluded from further analysis since their
spectrum can be well described by partial covering models (Church et
al. 1997; Sidoli et al. 2000), unlike the spectrum of 4U~0614$+$091.
Two of the other candidates (2S~0921$-$630 and 4U~1822$-$371) are
known accretion disk corona (ADC) sources.  These were also not
considered further due to distinct differences in their residuals from
the 0.7~keV feature we are concentrating on.  (The ADC sources have a
much broader residual, better described as a poorly fit continuum
rather than a line-like feature.)  
\centerline{\epsfig{file=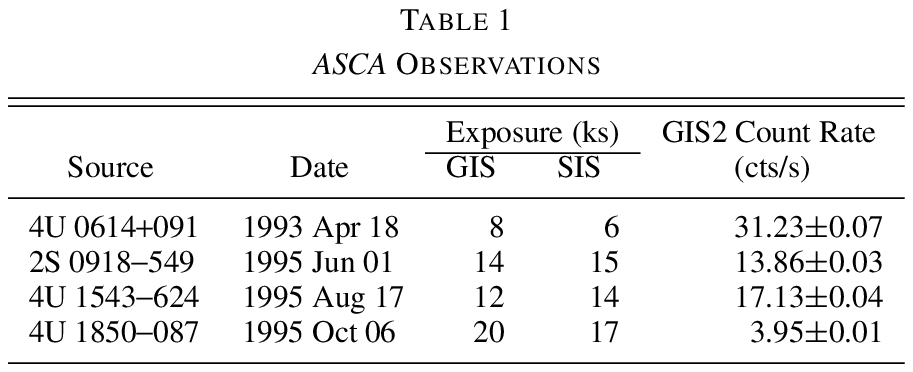,width=0.9\linewidth}}
\vspace*{0.02in}

\noindent
After narrowing down the candidates, we were left with four sources
with an almost identical feature at 0.7~keV.  These four remaining
sources include the three previously identified with a 0.7~keV
feature, as well as a new source: 4U~1850$-$087.

For these four sources, we made a detailed reanalysis of the relevant
data (see Table 1) obtained from the {\it ASCA\/} public data
archive, using the standard data reduction techniques recommended by 
the instrument team ({\it ASCA\/} Data Reduction Guide 2001).  We 
discarded telemetry-saturated SIS frames and applied any necessary deadtime
corrections to the GIS data.  Source and background spectra were
extracted from the screened event files, with $6^{\prime}$ and
$4^{\prime}$ radius source regions for GIS and SIS, respectively.  We used
the 2--10 keV GIS data to constrain the high energy continuum and the
0.4--7.0 keV SIS data to study the soft X-ray features.  

It was necessary to correct for pileup in the SIS data of 4U~0614$+$091, 
2S~0918$-$549, and 4U~1543$-$624.  Since the CCDs are only read every 
4--16 seconds, it is possible for sources brighter than 
$\sim$10~cts/s/sensor to have two or more photons detected at the same 
pixel.  When this happens, the instrument reads just one event at an 
energy comparable to the sum of the original photon energies, thus changing 
the shape of the spectrum.  The large point-spread function (PSF) of 
{\it ASCA\/} is useful in reducing the effect of pileup as well as 
correcting for it.  To correct for pileup, events in the inner core of 
the PSF should be rejected and the wings of the PSF used to predict the 
original spectrum of the source.  2S~0918$-$549 and 4U~1543$-$624 were both 
taken in 1-CCD mode and the standard pileup correction tool ({\tt corpileup} 
ftool) could be applied.  The observations of 4U~0614$+$091 were taken 
in 4-CCD mode so the standard correction tool could not be used.  Instead, 
we manually determined the area affected by pileup and rejected any events 
within this area.  Pileup correction was unnecessary for 4U~1850$-$087.  

We attempted to fit a variety of continuum models to these
spectra.  The simplest model that provided a reasonable fit was an
absorbed power-law $+$ blackbody model, although more complex models could
also be fit to the data.  In all cases, however, a strong residual
emission feature near 0.7~keV was present (see Figure~1, top panels).  These
residuals can be fit with a broad ($\approx$300 eV FWHM) Gaussian
emission line. Our fit results are summarized in Table~2.  The feature
is remarkably similar in all four sources, suggesting a common origin.
It is not seen in the {\it ASCA\/} data from other NS/LMXBs. 

\begin{figure*}[t]
\centerline{\epsfig{file=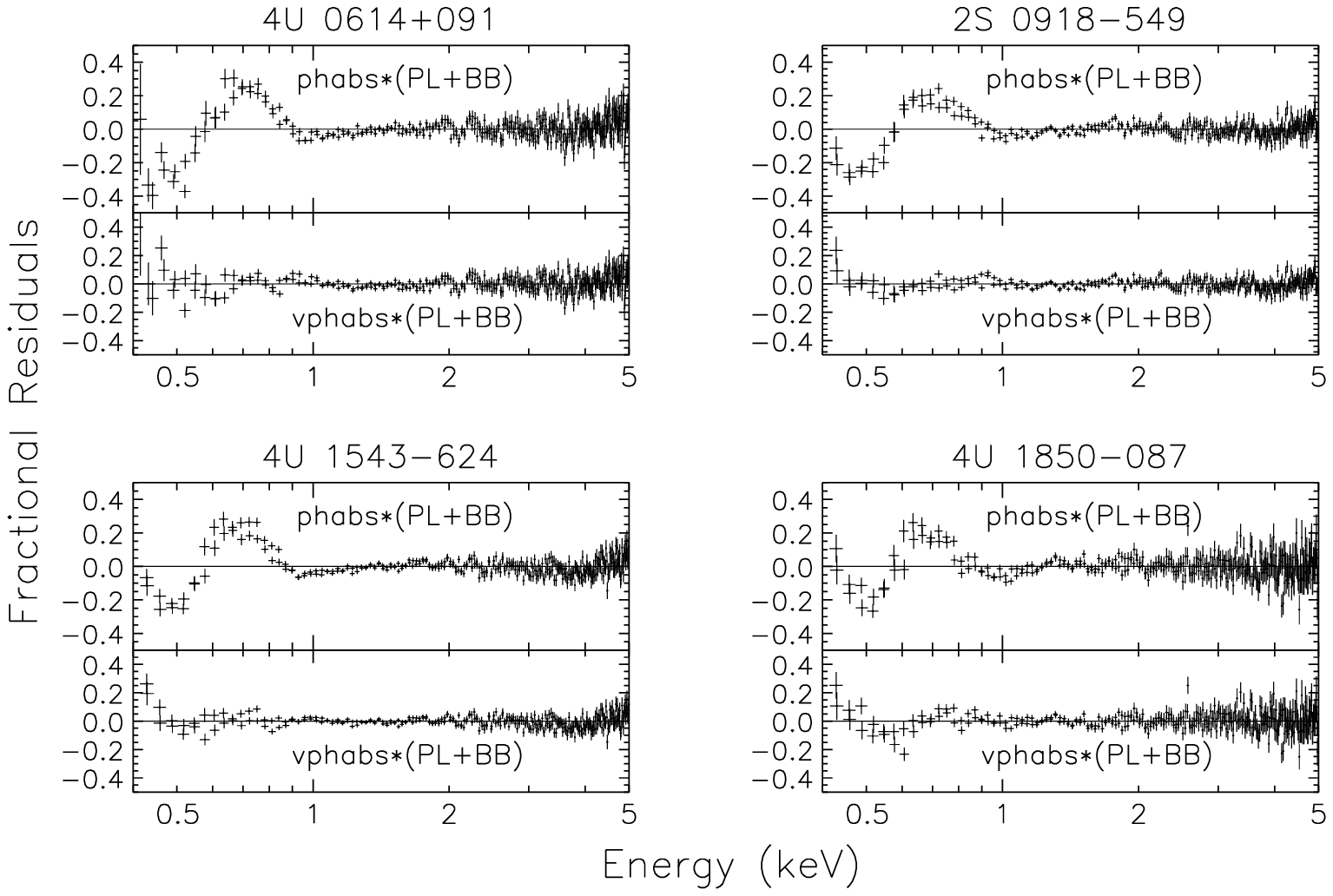}}
\caption{(upper panel) Fractional residuals ([data$-$model]/model) using 
a standard absorbed power-law $+$ blackbody model.  For clarity, only the 
SIS data are included.  (lower panel) Residuals of the same {\it ASCA\/} 
SIS spectra when a variable abundance absorption model (vphabs, with Ne and 
O allowed to vary) is used with the power-law $+$ blackbody model.}
\label{fig:1} 
\vspace*{0.05in}
\vspace*{\fill}
\centerline{\epsfig{file=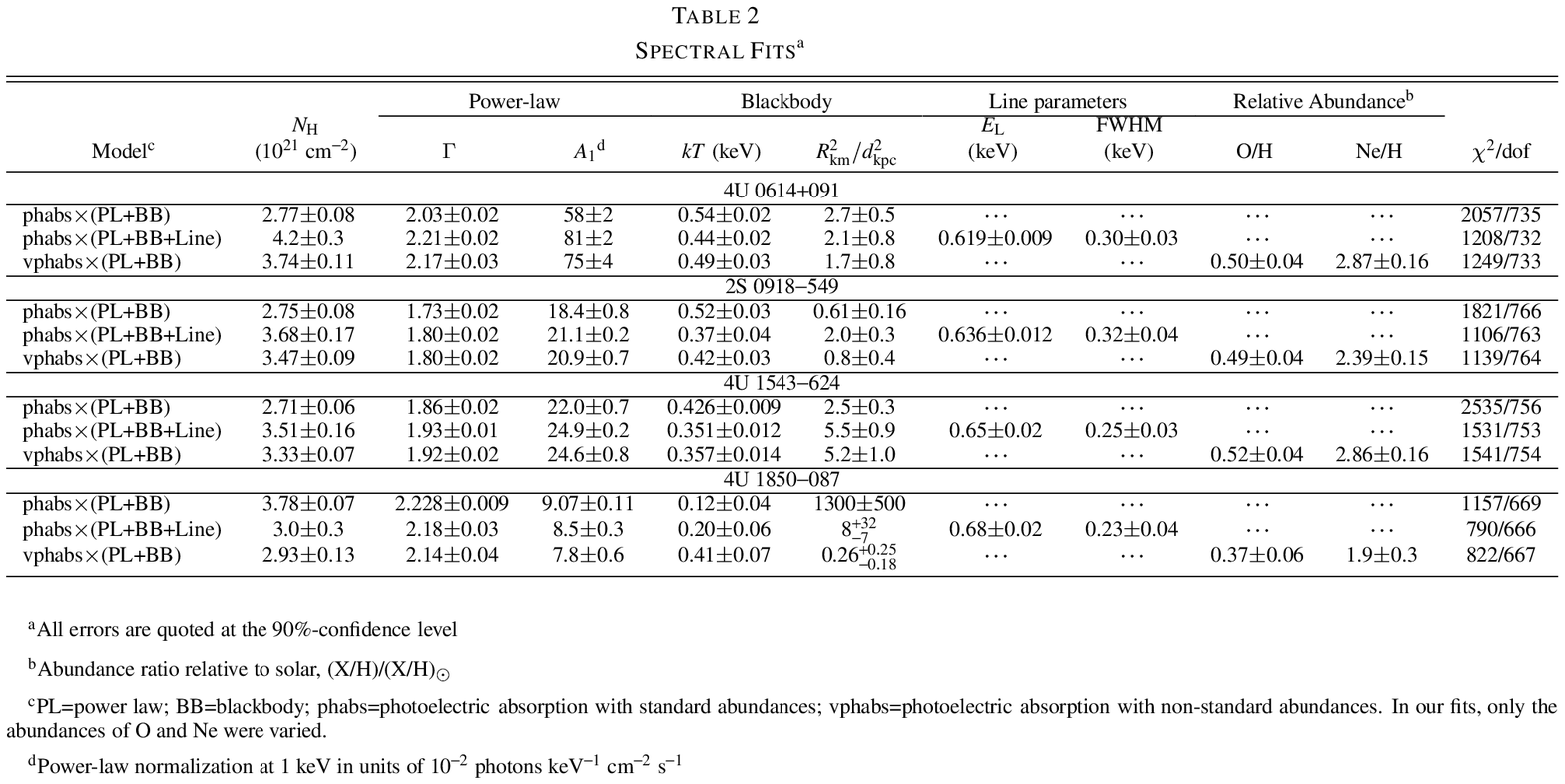}}
\end{figure*}

\section{An Alternative Model: High Neon Abundance}

Prior to the launch of {\em Chandra}, every previous observation of
the four sources identified above with sufficient statistics detected
the 0.7~keV feature.  The non-detection in 4U~0614+091 with the {\em
Chandra}/LETGS, with its superior resolution, is thus quite surprising
if the feature is indeed unresolved line emission.  Paerels et
al. (2001) attribute the non-detection to time variability of the
feature, although we note that this would be the only one of
$\approx$10 spectral observations of the source which did not detect
the feature.

We point out here, however, that it is possible to explain {\em all\/}
the previous data, including the {\em Chandra}/LETGS spectrum, without
recourse to time variability.  Paerels et al. (2001) found that the
relative strengths of the photoelectric absorption edges of O and Ne
in their spectrum of 4U~0614+091 could only be explained by an
anomalously large abundance ratio by number of Ne/O$\approx$1.25, a
factor of 7.5$\times$ larger than the solar value.  It is possible
that this is due to unusual abundances in the interstellar medium
along the line of sight.  Alternatively, there may be excess
absorption in neon-rich material local to the binary.  This would not
be unprecedented, since the presence of neon-rich local material has
been reliably established through both absorption edge and emission
line measurements in the LMXB pulsar 4U~1626$-$67 (Angelini et
al. 1995; Schulz et al. 2001).

Motivated by these points, we refit the {\em ASCA\/} data for these four
sources with a new absorbed power-law $+$ blackbody model, allowing the
relative abundances of O and Ne to vary (using the {\tt vphabs} model in 
XSPEC).  The best-fit parameters are given in Table~2, and the residuals 
from this model are shown in the bottom panels of Figure~1.   Note that a 
strong 0.7~keV feature is no longer present.   The remaining weak
structure in the low-energy residuals is consistent with previously
known, unexplained features in the SIS low energy response (Arnaud et
al. 2001).  As a consistency check, we note that our {\em ASCA\/} fit
results for 4U~0614+091 are in rough  agreement with the values
derived by Paerels et al. (2001) from their {\em Chandra}/LETGS
spectrum.

\section{Discussion}

We have shown that the ultracompact LMXB 4U~1850$-$087, like the
previously identified sources 4U~0614$+$091, 2S~0918$-$549, and
4U~1543$-$624, has a residual broad emission feature near 0.7 keV in
its {\em ASCA} spectrum when fit with an absorbed power-law $+$
blackbody model.  We have also shown that, in all four of these
sources,  the 0.7~keV feature is eliminated if the model includes a Ne
overabundance in the absorbing material along the line of sight.
This is consistent with the results of the recent high-resolution
{\it Chandra\/} observation  of 4U~0614$+$091, which found a stronger
than expected Ne absorption edge but no emission features (Paerels et
al. 2001).  

In principle, it is possible that the interstellar medium along all
four lines of sight is neon-rich, although we know of no plausible
mechanism for producing such a widespread enhancement.  Instead, we
conclude that there is cool material local to these binaries which is
neon-rich.  There is already strong evidence for neon-rich material
around another LMXB, the ultracompact ($P_{\rm orb}=42$ min) binary
pulsar 4U~1626$-$67 (Angelini et al. 1995; Schulz et al. 2001).  {\em
Chandra} observations of this source have resolved the individual
photoelectric absorption edges due to Ne and O, which are considerably
stronger than expected from the known value of $N_{\rm H}$ due to
interstellar material (Schulz et al. 2001). As noted earlier, this
source also shows a strong complex of Ne/O emission lines,
demonstrating a local origin for these elements.  Ultracompact
($P_{\rm orb}\lesssim 80$ min) binaries {\em must} have
H-depleted donors, usually white dwarfs (Nelson, Rappaport, \&
Joss 1986), and there are several indications that the mass donor in
4U~1626$-67$ is indeed probably a white dwarf (e.g., Chakrabarty
1998).  Schulz et al. (2001) show that either a C-O or an O-Ne white
dwarf can have a high Ne/O ratio, especially if crystallization has
occurred (see also Segretain et al. 1994, Gutierrez et al. 1996). 

Our four sources may also be ultracompact binaries.  Only 4U~1850$-$087
has a measured orbital period: $P_{\rm orb}=20.6$ min, making it
another ultracompact binary with a likely degenerate companion (Homer
et al. 1996).  The other three sources do not have known binary
periods.  However, their optical counterparts are faint and blue,
suggesting that the optical emission is dominated by the accretion
disk and that the donor has very low mass.  We can constrain their
orbital periods using the strong empirical correlation between
absolute magnitude $M_V$ and the quantity $\Sigma=(L_{\rm X}/L_{\rm
Edd})^{1/2}(P_{\rm orb}/{\rm 1\ hr})^{2/3}$ (van Paradijs \&
McClintock 1994), where $L_{\rm X}$ is the X-ray luminosity and
$L_{\rm Edd}$ is the Eddington luminosity.  Based on the available
measurements, we estimate that all four sources have $M_V$ in the
range 3.5--4.5.  The above correlation then gives $\log \Sigma\simeq
-1$, which for $L_{\rm X}\approx 10^{-2}\,L_{\rm Edd}$ implies that the 
binaries are indeed ultracompact with $P_{\rm orb}\lesssim$1~hr.
We therefore propose that 4U~0614$+$091, 2S~0918$-$549, 4U~1543$-$624, 
and 4U~1850$-$087 are all ultracompact binaries with low-mass, Ne-rich
degenerate dwarf donors and surrounded by neutral material, possibly
expelled from an accretion disk.

While the degenerate donor in an ultracompact binary must be
H-depleted, it need not necessarily be a C-O or O-Ne dwarf as in
4U~1626$-$67.  For example, the donor in the 11-min ultracompact LMXB
4U~1820$-$30 is known to be a He dwarf, on the basis of its X-ray
burst properties (Bildsten 1995).   However, we note that the apparent
{\em underabundance} of O in our four sources is not inconsistent 
with their having C-O or O-Ne donors.  Since the (unresolved) O
absorption edge dominates the {\em ASCA} determination of $N_{\rm H}$,
the true $N_{\rm H}$ values would be considerably lower if there is a
large amount of O-rich material local to the binaries.  A direct
Ly$\alpha$ measurement of $N_{\rm H}$ might clarify this point, as has
already been  done for 4U~1626$-$67 (Wang \& Chakrabarty 2001; Schulz
et al. 2001).  Another way to determine the donor composition is
through emission line spectroscopy.  No reliable X-ray lines have been
detected for any of the sources.  In the optical spectrum of
4U~0614+091, no H or He lines are seen, although there is a weak
detection of the \ion{C}{3}/\ion{N}{3} Bowen emission blend near
4640~\AA\ (Davidsen et al. 1974; Machin et al. 1990).  This is
essentially identical to the optical spectrum of 4U~1626$-$67 (Cowley,
Hutchings, \& Crampton 1988; Wang \& Chakrabarty 2001). 

One crucial difference between our four binaries and 4U~1626$-$67 is
that the latter is a pulsar with a strong ($3\times 10^{12}$~G;
Orlandini et al. 1998) magnetic field, while our four sources are
presumed to contain weakly magnetized neutron stars.  This may result
in a different ionization structure in the disk, explaining the
absence of a strong Ne line complex in the weak-field systems.
These systems should also be subject to thermonuclear X-ray bursts
(Joss \& Li 1980), although unusual donor composition might lead to
atypical burst properties.  In particular, calculations for
thermonuclear flashes in pure C layers indicate very long ($\gtrsim
0.5$ yr) recurrence times and correspondingly large burst fluences; 
these general trends will be true even for C mass fractions as low as
10\% (Cumming \& Bildsten 2001), which is relevant for C-O donors.  
Few type~I X-ray bursts have been reported from our four sources:
two bursts from 4U~0614+091 (Swank et al. 1978; Brandt et al. 1992),
one from 2S~0918$-$549 (Jonker et al. 2001), and three from
4U~1850$-$087 (Hoffman, Cominsky, \& Lewin 1980).  However, none of
these bursts has an unusually large fluence, and two of the bursts
from 4U~1850$-$087 were separated by only 17~hr.  This may indicate
that the donors in these systems are not C-O dwarfs.  In that case,
some other mechanism for producing a Ne enhancement is required.

\acknowledgements{We thank Lars Bildsten and Norbert Schulz for useful
discussions.  This work made use of data from the High Energy
Astrophysics Science Archive Research Center (HEASARC) at NASA Goddard
Space Flight Center, and was supported in part by NASA under grant
NAG5-9184 and contract NAS8-38249.}

\end{document}